\documentclass[fleqn,10pt]{wlscirep}

\usepackage{color}
\usepackage[utf8]{inputenc}
\usepackage[T1]{fontenc}
\usepackage{bbm}  
\usepackage{graphicx}
\PassOptionsToPackage{hyphens}{url}
\usepackage{hyperref}
\usepackage[T1]{fontenc}
\usepackage[normalem]{ulem} 
\usepackage{soul}

\title{Impact of urban structure on infectious disease spreading}

\author[1]{Javier Aguilar}
\author[2,3]{Aleix Bassolas}
\author[4,5]{Gourab Ghoshal}
\author[4]{Surendra Hazarie}
\author[6]{Alec Kirkley}
\author[1,7]{Mattia Mazzoli}
\author[1]{Sandro Meloni}
\author[4]{Sayat Mimar}
\author[2]{Vincenzo Nicosia}
\author[1,*]{Jos\'e J. Ramasco}
\author[8]{Adam Sadilek}

\affil[1]{Instituto de F\'{\i}sica Interdisciplinar y Sistemas Complejos IFISC (CSIC-UIB), Campus UIB, 07122 Palma de Mallorca, Spain}
\affil[2]{School of Mathematical Sciences, Queen Mary University of London, E1 4NS, London, United Kingdom}
\affil[3]{Departament d'Enginyeria Informatica i Matematiques, Universitat Rovira i Virgili, 43007 Tarragona, Spain}
\affil[4]{Department of Physics \& Astronomy, University of Rochester, Rochester, NY, 14627, USA}
\affil[5]{Department of Computer Science, University of Rochester, Rochester, NY, 14627, USA}
\affil[6]{School of Data Science, City University of Hong Kong, 999077, Hong Kong}
\affil[7]{INSERM, Sorbonne Universit\'e, Institut Pierre Louis d'Epid\'emiologie et de Sant\'e Publique, IPLESP, Paris, France}
\affil[8]{Google, 1600 Amphitheatre Parkway, Mountain View, CA, 94043, USA}

\affil[*]{jramasco@ifisc.uib-csic.es}

\begin{abstract}
The ongoing SARS-CoV-2 pandemic has been holding the world hostage for
more than a year now. Mobility is key to viral spreading and its restriction is the main non-pharmaceutical interventions to fight the virus expansion. Previous works
have shown a connection between the structural organization
of cities and the movement patterns of their residents. This puts
urban centers in the focus of epidemic surveillance and interventions.
Here we show that the organization of urban flows has a tremendous
impact on disease spreading and on the amenability of different
mitigation strategies.  By studying anonymous and aggregated
intra-urban flows in a variety of cities in the United States and
other countries, and a combination of empirical analysis and
analytical methods, we demonstrate that the response of cities to
epidemic spreading can be roughly classified in two major types
according to the overall organization of those flows. Hierarchical
cities, where flows are concentrated primarily between mobility
hotspots, are particularly vulnerable to the rapid spread of
epidemics. Nevertheless, mobility restrictions in such types of cities
are very effective in mitigating the spread of a virus. Conversely, in
sprawled cities which present many centers of activity, the spread of
an epidemic is much slower, but the response to mobility restrictions
is much weaker and less effective. Investing resources on early
monitoring and prompt ad-hoc interventions in more vulnerable cities
may prove helpful in containing and reducing the impact of future pandemics.
\end{abstract}

\begin{document}

\flushbottom
\maketitle

\thispagestyle{empty}

\section*{Introduction}

The SARS-CoV-2 virus, believed to have originated in the Hubei
province of China around the end of 2019~\cite{Zhou_2020,Wu_2020}, is still active while presenting new threats in terms of new variants of concern. Given the slow pace of vaccination, especially even though not only in developing countries, the response toolbox of public health authorities worldwide has strongly relied on non-pharmaceutical interventions, such as social distancing, travel bans and stay-at-home orders~\cite{fang2020,haug2020,pepe2020,chinazzi2020,dehning2020,gatto2020,kraemer2020,maier2020,melo2021,pan2020,schlosser2020,oh2021mobility}. 

Human mobility is known to play a central role in the spreading process ~\cite{flahault1992,grais2003,hufnagel2004,brownstein2006,colizza2006,colizza2007,tatem2012,finger2016,zhang2017,barbosa2018, massaro2019} as it was the case for the first wave that hit China following the inter-city mobility from Wuhan \cite{kraemer2020,mu2020,niu2020}. In a connected planet, rapid world-wide spread is enabled by long-distance air-, land- and sea-transportation among countries and continents, which subsequently expand the disease to new and susceptible area of the world ~\cite{brownstein2006,colizza2007,balcan2009, gilbert2020,kraemer2020, rader2020, mu2020, niu2020,mazzoli2021interplay,kraemer2021spatiotemporal}. While early travel restrictions contribute to delayed disease spread, their utility is much reduced if the disease has incubation periods longer than a day or if there is asymptomatic transmission~\cite{ferguson2005,hollingsworth2006,epstein2007,bajardi2011,meloni2011,poletto2014,fang2020, pepe2020,chinazzi2020,pan2020,gatto2020,pei2020}. Long-range mobility is mainly driven by air transportation, but restrictions to international flights have shown limited utility in mitigating the propagation of infectious diseases unless they are applied very early and in a comprehensive manner~\cite{bajardi2011,poletto2014,chinazzi2020}. 

The majority of extant work on the role of mobility in regulating the spread of pandemics, has focused on long range trips between countries, states and cities. Once restrictions on land, sea and air travel are imposed, and long range transportation networks cease to play a role, the core of contagion keeps developing in the main center of human activity: the densest and most crowded cities of the globe. In urban areas, transportation modes include vehicular traffic, pedestrian, bike, and public transportation modes such as metros and buses, are predominant~\cite{varga2016,barbosa2018}. In particular, the mixture of different transportation modes at an urban level coupled with 
the spatial organization of socioeconomic activities in residential areas, creates
unique mobility fingerprints~\cite{Lee_2017, Kirkley_2018, soriano2018,bassolas2021diffusion,gauvin2021socio,heroy2021covid,valdano2021} in which crowded stations, offices and stores are continuously packed and close contact is inevitable. Historically, epidemics have strongly influenced planning of cities~\cite{banai2020}. In the history of urban planning, e.g. after the black plague or cholera, ventilation of streets, indoor daylight, large public squares and parks, were all architectural recipes realized in order to mitigate the spread of disease, in particular respiratory ailments such as tuberculosis. Such measures helped reduce the density of cities, created safer and less infectious residential areas and increased the opportunities for physical distancing in public spaces, while allowed for better isolation of affected individuals from the larger urban population \cite{de2000fear,eltarabily2020post,martinez2021}.

 Currently more than half the world's population lives in urbanized areas, while forecasts see this number growing to $68\%$ in 2050 \cite{unurban}. A vigorous debate on how to redesign urban spaces to respond to future challenges stemming from the post-pandemic era has emerged~\cite{grant2020cities,eltarabily2020post}. While there has been prior work on the role of mobility on viral spread in urban systems \cite{mu2020, niu2020,rader2020}, showing how dense urban areas exhibit more diffuse epidemics of influenza \cite{dalziel2018} and higher attack rate peaks of SARS-Cov2 \cite{rader2020}, scarce attention has been devoted to the intrinsic spatial structure of intra-urban mobility, and its effect on the evolution of urban outbreaks. The way in which human mobility affects an epidemic progression is by regulating the population mixing patterns. This is not only an important factor at international and national levels, but it is critical in urban environments, where most close contacts happen, particularly in the case of airborne diseases such as COVID-19 or influenza, given that contagions may occur at fortuitous encounters due to shared transportation media, crowded areas, and business activities. Hence, understanding how the spreading of a disease depends on the structural organization of a city mobility and its attendant population mixing patterns, is of fundamental importance to devise adequate and tailored containment strategies and surveillance protocols \cite{lee2020epidemic}. 

 Here we investigate how the spatial structure of urban mobility shapes critical features of the local spreading patterns and the efficiency of the potential mitigation measures. We showed in a previous analysis that when it comes to the internal organization of city-wide mobility patterns, urban areas can be classified in a spectrum between centralized-hierarchical and
decentralized-sprawled~\cite{bassolas2019}. The level of
centralization can be encapsulated in a metric $\Phi$, called the
flow-hierarchy. As a first step, we analyze official surveillance data at county scale in the US and human mobility records from the Google COVID-19 Aggregated Mobility Research Dataset. The analysis performed on 22 of the most populated cities of the US indicates that intra-urban mobility hierarchy is connected to higher incidence peaks and faster epidemic growths in cities. We confirm the empirical observations through a compartmental model of epidemic spread on cities with different levels of $\Phi$ showing higher peaks and faster outbreaks in more hierarchical cities as compared to sprawled ones, independent of their population density or shape. Moreover, our formulation enables analysis of "what-if" scenarios to assess the effect of mitigation policies, demonstrating that hierarchical cities perform and respond better at containing the epidemics when lockdowns are issued. The model is finally performed repeatedly for higher and lower values of $R_0$ in order to check for the stability of results in the context of new variants or other diseases, which can show higher or lower infectiousness. Our results are of fundamental interest in the context of global and national surveillance systems, improving risk-perception of epidemics, and in order to prevent new waves of infections caused by new COVID-19 variants and new viruses in general.

\section*{Methods} 

\subsection*{Mobility data}

The mobility flows are sourced from the
Google COVID-19 Aggregated Mobility Research Dataset, containing anonymized trips weekly aggregated over users who have turned on the Location History setting, which is off by default. The dataset aggregates flows of populations between S2
cells (\url{https://github.com/google/s2geometry}) of approximately 5 km$^2$. To produce this dataset, machine learning is
applied to logs data to automatically segment it into semantic trips~\cite{bassolas2019}. To provide strong privacy guarantees, all trips were anonymized and aggregated over one week using differential privacy~\cite{48778}. The research presented here is carried out on the resulting weekly flow data. No individual user data was ever manually inspected, only heavily aggregated flows of large populations were
handled. To further confirm the model results, we have additionally used commuting data at zip code level from the Longitudinal Employer-Household Dynamics surveys of the US census office. In particular, the information comes from the LODES dataset concerning commuting in 2016-2017 (see Data Data availability statement for the download link).

\subsection*{Urban mobility and its hierarchy}

As mentioned, the space of the urban areas is divided in S2 cells. The weekly trip flows between pairs of areas are incorporated in an Origin-Destination matrix $\mathbf{T}(t)$, whose elements $T_{ij}(t)$ encode the trip flow at week
$t$ between the two spatial units $i$ and $j$. The diagonal terms $T_{ii}$ correspond to movements within the area $i$.  The total flow of a territory is the sum of the trips over all the S2 cells present in the city, $T = \sum_{i,j} T_{ij}$. In general, the relative trip-flow change between two timestamps $t_1$ and $t_2$ is calculated as $\left(T(t_1)-T(t_2)\right)/T(t_2)$. Note that when we estimate the reduction after lockdown, $t_2$ refers to the mobility prior to the restrictions and $t_1$ after the lockdown. 

To calculate $\Phi$, a hotspot level must be assigned to each geographical unit as explained in Ref.~\cite{bassolas2019}. In a nutshell, a Lorenz curve is constructed from the trip outflow of each area; the curve contains information on the differences in out-mobility between cells by ordering them in increasing order in terms of outflow and representing the fraction of cells versus the fraction of total outflow. Taking the derivative at $(1,1)$ establishes a threshold. Those cells over the threshold are classified as level-1 hotspots~\cite{louail2014}, and the process is applied iteratively with remaining cells until all of them have an assigned level.  Once every cell $i$ is assigned a level $L_i$, trip flows between cells are aggregated in a matrix $S_{\ell m}$ as: 
\begin{equation}
S_{\ell m} = \frac{\sum_{i< j} T_{ij} \, \delta(L_i,\ell) \, \delta(L_j,m)}{\sum_{i < j} T_{ij}}
\end{equation}
where $\ell$ and $m$ are levels and $\delta(x,y)$ is the Kronecker
delta. The flows $S_{\ell m}$ are normalized by the
total flow in the city. Self-flows within cells are not considered
when calculating $S_{\ell m}$. With these definitions, the flow
hierarchy $\Phi$ is calculated according to~\cite{bassolas2019}:
\begin{equation}
\Phi = \sum_{\ell,m} S_{\ell m} \, \left\{ \delta(\ell,m) + \delta(\ell,m-1) + \delta(\ell-1,m)\right\} .
\end{equation}
The value lies in the range $0 \leq \Phi \leq 1$ with the limiting cases corresponding to flows being distributed to levels far from the considered hotspots, or flows only being concentrated at similar levels. In hierarchical cities, trips are concentrated between mobility
hotspots at higher levels (spatially clustered), with $\Phi$ taking on values closer to one. In sprawled cities, hotspots display a homogeneous spatial distribution and the flows are distributed more evenly across levels, with $\Phi$ taking on a lower value. Empirically measured values for global cities typically lie in the range $0.7 \leq \Phi \leq 0.95$. While the range appears superficially narrow, the differences track well with significant variability in a variety of urban indicators associated with different levels of public transportation usage, walkability, health indicators and levels of pollution~\cite{bassolas2019}. The metric is also well correlated with considerably more complex composite measures (integrating socioeconomics, land-use, population density among others) such as so-called sprawl indices~\cite{ewing2015}, and outperforms them as a predictor of urban indicators.

\subsection*{Epidemic data}

 The epidemic data has been downloaded from two sources (USAFacts and NY Times) listed in the Data availability statement below. Despite, the fact that we have global mobility data, in what is to follow, we focus on cities in the United States. This is done primarily for two reasons: first, US cities show a wide spectrum of configurations from hierarchical to extended or sprawled, with New York city and Atlanta as archetypal examples. Second, due to inherent biases and noise in collection of epidemic data, it is problematic to compare between different countries. We also restrict our analysis to the time-period corresponding to the so-called first wave of infections. For the purposes of our analysis we collected data on the 22 largest metropolitan areas in the United States, in terms of the number of counties and with populations exceeding 2 million inhabitants. The full list of cities and their corresponding values of $\Phi$ are shown in SI Table~S1. The data resolution is at the level of counties, which in a large city are well populated units with reasonable statistics (SI Tables S2-S23 detail the list of counties considered per OECD urban area).
 
\subsection*{Metrics of epidemic severity}

The epidemic evolution is studied using the incidence curve as basis, which corresponds to the number of new cases detected in a day per capita. In the case of the data, the number of detected individuals depends on the testing policy, while in the models these are all the new infected individuals in each simulated day. The incidence can then be represented by curves $I_j^u(t)$ for each county $j$ of city $u$ at day $t$. These incidence curves can be aggregated at full city scale by summing the cases in the counties forming the urban area and dividing by the total population, $I^u(t)$. The first metric will be the maximum value of the curve at city $u$, $I^u_{max}$. In the case of the empirical data and to avoid fluctuations, we use a 14 days aggregation of the incidence before the peak. 

Besides the direct height of the incidence peak, we are interested in how quick is the spreading process initially in every population. To quantify this, we calculate $R_{\rm eff}$ that is the effective reproduction number. This has been done using the methodology developed in \cite{bettencourt2008real} and the code available at \cite{tizzoni2020}. This method relies on a Bayesian dynamic inference of the reproductive number based on a stochastic epidemic model informed by an estimation of the serial interval mean. The meaning of $R_{\rm eff}$ is related to the exponential change of the incidence: the curve grows if $R_{\textrm{eff}} > 1$, and it decays if $R_{\textrm{eff}} < 1$. $R_{\textrm{eff}}$ can be measured at different geographical scales but it has a more practical meaning in localized regions such as cities or counties where it can accurately reflect the contagion patterns. We then define $R_{early}$ as the average over three weeks of the effective reproductive number measured at the early epidemic stages (after the city incidence has gone over a certain case onset: 100 for the empirical data and 10,000 for the models where there is no detection problem). For a robustness test, we applied the same methodology on other types of data from the UK comparing $R_{\textrm{eff}}$ and the $R_{early}$ computed on the reported cases by publication date with cases by specimen date, finding strongly consistent results (see SI Figs. S34 and S35). 
Finally, in order to define the counties synchronization in the epidemic evolution, after mobility restrictions are implemented in a city we define $R_{p,response}$ as the Pearson correlation coefficient of the consecutive five weeks $R_{\rm eff}$ and mobility reduction of each area from the onset of the first 100 registered cases.

\subsection*{Model description}

The simplest version of our model is structured in a meta-population framework with Susceptible (S), Exposed (E), Infected (I) and Recovered (R) compartments and taking as basic spatial units the S2 cells provided by the mobility data. For a given city, the population $P$ is distributed among the S2 cells $i$ according to their mobility, $p_i = P \, (\sum_j T_{ij})/T$ where the index $j$ runs over all cells including $i$ and $T$ corresponds to the total trips of the city. Note that while, this is not a realistic configuration, the goal here for the model is to uncover the effect of mobility hierarchy, and not necessarily to forecast epidemic evolution, which would require more precise data on the population mixing. Usually, only a fraction of people, $M$, leave the residence area, as for instance in the case of commuting between zip codes, a value of $M \approx 0.4$ was measured in US cities (see SI Figs. S6 and S7). A similar value was observed in previous works for commuting in more extensive areas across the world~\cite{balcan2009,balcan2010,oh2021mobility}. Unless otherwise specified, the particular parameters selected for the model are shown in Table \ref{table1}.
 
 \begin{table}[h!]
\begin{centering}
\begin{tabular}{|c|c|}
\hline 
\hline
Parameter & Explanation\\
\hline
$\beta=0.48$ $\textrm{days}^{-1}$  & Infection rate$\left(S+I\stackrel{\beta}{\rightarrow}E+I\right)$.\tabularnewline
\hline 
$t_{I}=3.8$ $\textrm{days}$ & Average infectious time$\left(I\stackrel{t_{I}^{-1}}{\rightarrow}R\right)$.\tabularnewline
\hline 
$t_{E}=3.7$ $\textrm{days}$ & Average exposed time$\left(E\stackrel{t_{E}^{-1}}{\rightarrow}I\right)$.\tabularnewline
\hline 
\hline
\end{tabular}
\par\end{centering}
\caption{Epidemic model parameters obtained from Ref. \cite{domenico2020}.  \label{table1}}

\end{table}

\begin{figure}
\begin{center}
\includegraphics[width=8cm]{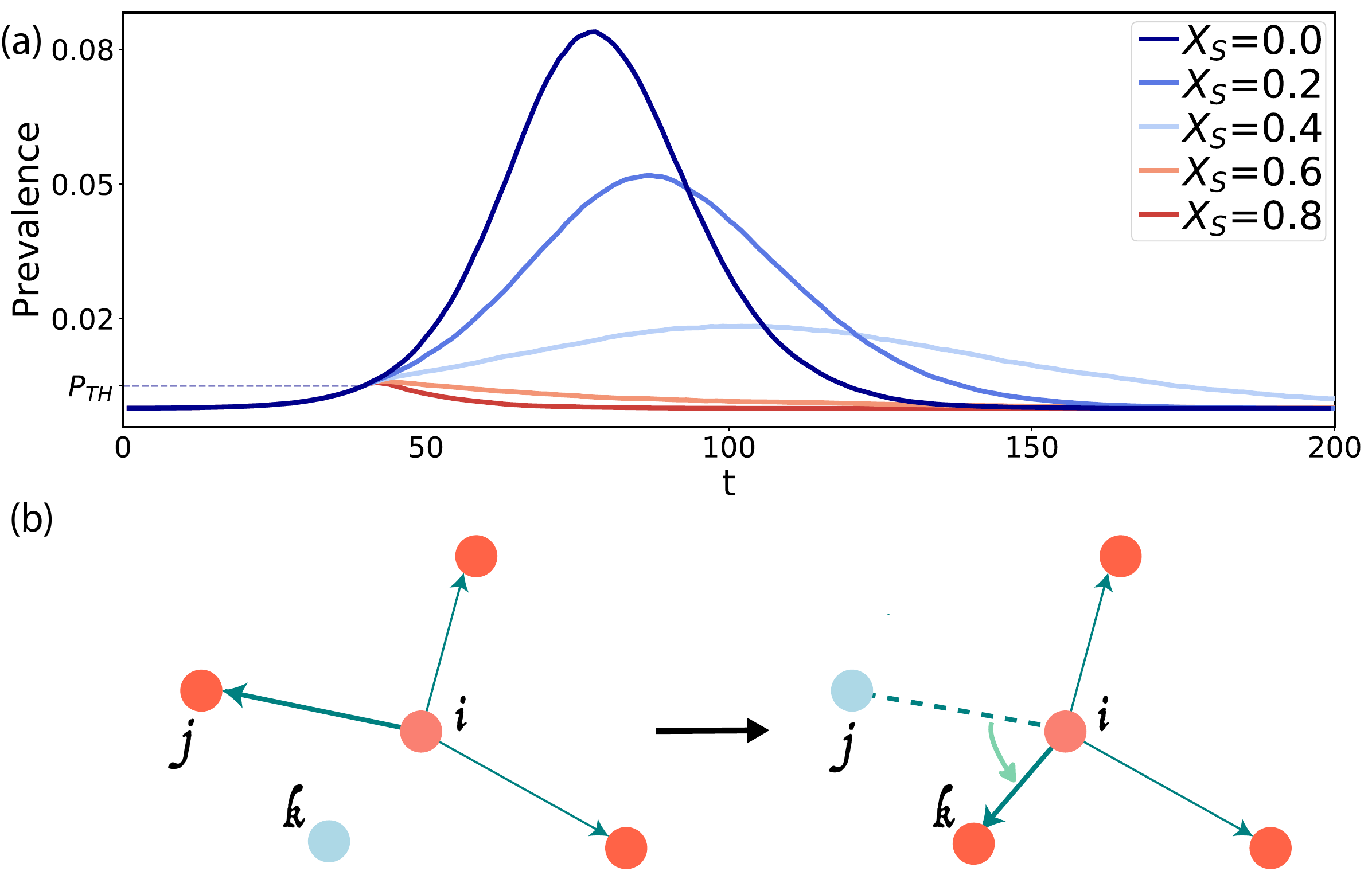}
\caption{Modeling disease spread in a city.  In (a), representative prevalence curves for different values of $X_S$, the fraction of susceptible individuals that do not participate in the infection process, mimicking the effect of isolation and stay-at-home orders. The curves show the case for the mobility in Atlanta at the resolution of S2 cells as geographical units. In (b), a sketch showing the reshuffling procedure to obtain different values of $\Phi$.
  \label{fig:model1}}
\end{center}
\end{figure}

Consequently in our model, when there are no travel restrictions, a fraction $M = 0.4$ of the population of every cell $i$ moves and selects their destination proportional to outflow $T_{ij}$ from $i$ to the other cells $j$ (including $T_{ii}$). Non-moving individuals are not excluded from the epidemic dynamics but contacts occur inside their cell of residence. Note that we assume that the flows represent recurrent mobility, hence individuals return home after a time $\tau$ set at 8 hours. The incorporation of recurrent mobility in epidemic dynamics is done by using a classical approach that  encodes the information of the mobility matrices into an effective transition rate~\cite{sattespiel1995,balcan2009,balcan2010}, see SI section S8 for details. Cities in this initial idealized configuration maintain their empirical $\Phi$ because the outflows are proportional to those in the empirical network regardless of the value of $M$. 

The main outcome of the model are the epidemic curves for each geographical unit (S2 cells). These cells have to be aggregated to county or city levels to allow for a fair comparison with the results obtained with the epidemiological data. A sketch with the different geographical scales involved and the type of data available at each of them can be found in SI Fig. S37.

\subsection*{Inclusion of lockdowns} 

To model the effect of lockdowns on mobility, we need to consider that mobility flows change in both magnitude and destination in response to restrictions, with reductions that can reach up to $80\%$ of the flows compared to baseline levels. For the purposes of simplicity and tractability, for each city we employ two mobility networks: one considering a typical day prior to restrictions with elements $T_{ij}$ as used thus far, and another, after mobility restrictions, with a modified mobility matrix $T'_{ij}$. The total trips per cell $i$ for each of the scenarios are then  $T_i = \sum_{j} T_{ij}$ and $T'_i = \sum_{j} T'_{ij}$.  We propose $M^{'}_i= M \, (T'_i/T_i)$ as the new fraction of individuals traveling outside of their residence under mobility restrictions. Since travel flows in matrix $\mathbf{T'}$ are usually smaller than flows in $\mathbf{T}$, this results in an overall reduction of mobility.

During stay-at-home orders, part of the population remained isolated or only interacted with individuals in their household; not participating in the global spreading process directly. This effect has been modeled following the literature~\cite{domenico2020,aleta2020,arenas2020}, where a fraction $X_S$ of susceptible individuals do not participate in the infection process. To model that restrictive measures are enforced when outbreaks are detected, we assume that restrictions are imposed once the prevalence (the total number of active infected cases per day) reaches a certain threshold value $\pi_{\textrm{th}}$. In Fig. \ref{fig:model1}~(a), we plot some representative examples for the prevalence $\pi$ for a variety of mitigation scenarios governed by varying $X_S$. Stronger lockdowns flatten the curves and can even suppress propagation as occurs for $X_S = 0.9$. 

\subsection*{Generating instances of a city with different $\Phi$}

To produce instances of a single city with different values of $\Phi$, we reshuffle a portion of randomly selected links, for instance the one going from area $i$ to $j$, $T_{ij}$, by picking a new destination at random ($k$) while maintaining the link-weights. In case a link already existed between the original area $i$ and $k$, the weights of the links $i \to j$ and $i \to k$ are interchanged. This procedure creates multiple realizations of the city in terms of the arrangement of mobility flow, while preserving the population distribution and densities, as well as the total number of trips along with the number of destinations per origin. The procedure is applied equally to prior and post confinement scenarios, selecting the same new random destination for every pair $T_{ij}$ and $T'_{ij}$ (for a sketch see Fig. \ref{fig:model1}~(b)). 

\section*{Results}

\subsection*{Empirical local propagation patterns and city organization}

\begin{figure*}
\begin{center}
\includegraphics[width=14cm]{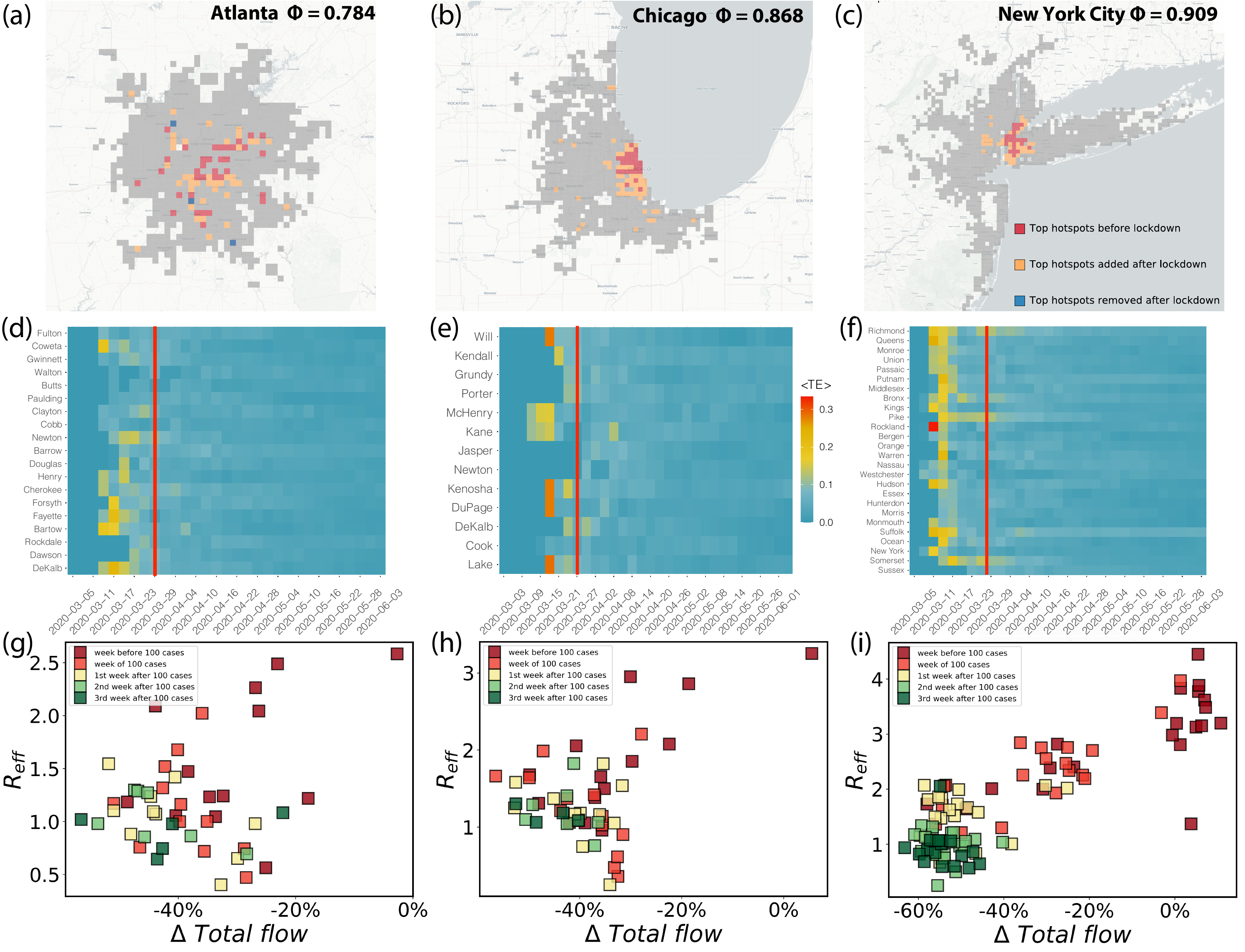}
\caption{Types of cities and COVID-19 spreading. Maps
  with the changes in mobility hotspots before and after the
  lockdown in three cities with different mobility hierarchy (higher $\Phi$ indicates more hierarchical cities): (a-c) 
  Atlanta, Chicago and New York City, respectively, in the week of February 2 for pre-lockdown mobility and the week April 5 for the post-lockdown. (d-f) The average Transfer Entropy $\langle TE \rangle$, capturing the influence of an administrative division (county) to drive infection-spread as a function of time.  Vertical red
  lines mark the date of the official lockdown. After lockdown, the ability of a single region to drive infection spread dissipates, and the transmission evolves independently in each area. (g-i) The temporal evolution of the effective reproduction number before and after lockdown versus the mobility change one week before $R_{\textrm{eff}}$ is measured. Each symbol represents a county of the city. While sprawled cities like Atlanta have regions responding independently, in centralized New York City, we see a clear synchronized and monotonically decreasing reduction in $R_{\textrm{eff}}$ as a function of mobility reduction.}
\label{fig:transfer}
\end{center}
\end{figure*}

The primary response to the early onset of the epidemic were mobility restrictions corresponding to stay-at-home measures, modifying the spatial patterns of urban mobility and disrupting most of the routes used by the virus to propagate. In Fig.~\ref{fig:transfer}~(a-c), we show the observed spatial mobility layout for three cities arranged in increasing levels of centralization in their mobility structure: Atlanta ($\Phi=0.784$), Chicago ($\Phi =0.868$) and New York City NYC ($\Phi =0.909$). In each case, shown as red dots are the top mobility hotspots prior to lockdown. In the case of Atlanta, being a decentralized city, we see that the mobility hotspots were spread across the urban area, whereas Chicago and NYC being more hierarchical had their hotspots clustered in a centralized fashion. The new mobility hotspots that emerged after mitigation measures are shown as orange dots, whereas in blue, we show those areas that ceased to be top hotspots. For hierarchical cities such as NYC or Chicago, while the original top hotspots remained in the reduced mobility phase, new areas of mobility activity emerged in an extended fashion, effectively distributing the mobility over larger parts of the city, thus reducing agglomeration. 
This feature was observed across several cities in the United States, as reflected by plotting in decreasing order, the fractional change in distance between hotspots due to mobility restrictions as shown in SI Fig. S1.

The resolution of the epidemic data allows us to run a granular analysis of the impact of lockdowns by investigating the extent to which specific areas of the city drove the epidemic spreading to other parts. To do so, we calculated the average Transfer Entropy, $\langle \mathrm{TE} \rangle$, between the incidence curves for each subsets of the city (see SI Section~2 for details), and in Fig.~\ref{fig:transfer}~(d-f) we plot the results as a matrix whose rows correspond to the administrative sub-unit, and columns correspond to weeks, starting from the onset of 100 cases in the whole city, to the first week of June. The elements in the matrix are color-coded by the value of $\langle \mathrm{TE} \rangle$. For all three cities, a few regions drove the spreading of the infection before lockdown, although the strength of the driving was stronger on average in NYC and Chicago compared to Atlanta (results for four other cities, including three outside the US are reported in SI Fig.~S2). Once lockdown started (marked as vertical red lines), the driving became diffused, indicating the predominance of localized spreading in sub-regions with little influence on one another. The localization in spreading appeared in parallel to the equivalent phenomenon in mobility, with a relative increase in self-flows and a decrease of inter-area flows in the administrative units as shown in SI Fig. S3. While the influence of each administrative unit on infection-spreading dissipated in a similar fashion, there is a key difference in how they were synchronized in terms of their response to mobility-mitigation, as we show next.

\begin{figure}
\begin{center}
\includegraphics[width=10cm]{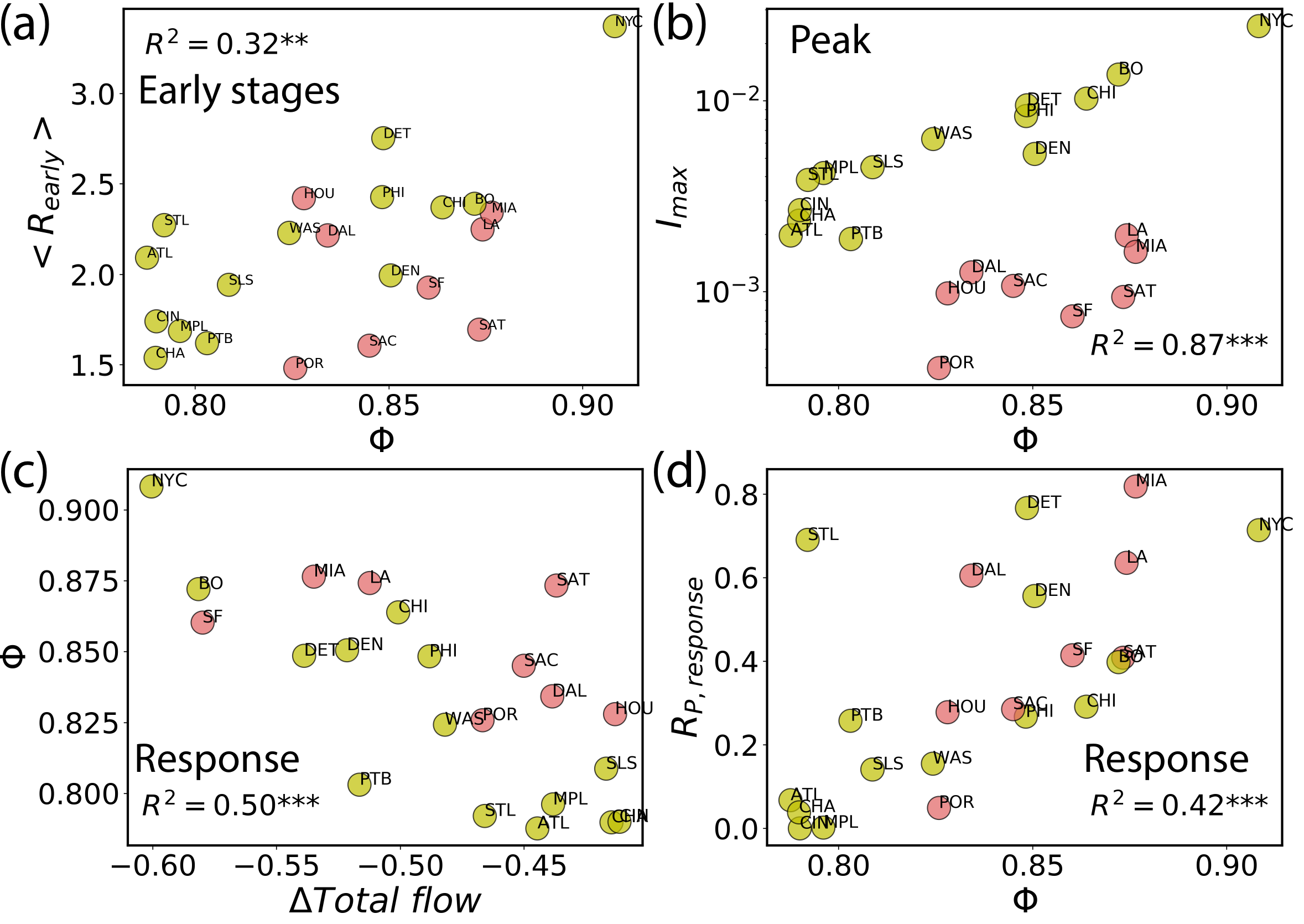}
\caption{Connecting hierarchy with epidemic features and mitigation efforts. Shown are the 22 cities in the United States, as of June 14, 2020 in terms of the pandemic situation. Cities in pale yellow have already peaked, while infections continue to grow in those marked in red. The figure suggests the extent of spread is strongly correlated with centralization. (a)
  Average observed $R_{\textrm{eff}}$ over three weeks after the
  onset of 100 cases as a function of $\Phi$, namely $R_{early}$. Initial observed transmission increased with centralization. (b) Accumulated number of reported new cases per capita two weeks before the maximum incidence $I_{\textrm{max}}$. In
  (c), $\Phi$ versus observed relative decrease in total flow. Mobility reductions were much more drastic in hierarchical cities. (d) Synchronization of mobility reduction and contagion spread among city counties measured through the Pearson coefficient of plots as those shown in Fig. \ref{fig:transfer} (g-i) for Atlanta, Chicago and NYC. Response to mitigation was more sensitive in cities with higher $\Phi$.}
  \label{fig:cities}
\end{center}
\end{figure}

In Fig.~\ref{fig:transfer}~(g-i), we plot the quantity $R_{\rm eff}$ as a function of mobility reduction in each city with points corresponding to sub-units and colored by time-period starting from the first 100 registered cases. Note that neighboring counties could have a correlation in $R_{\textrm{eff}}$ due to strong mobility and imported cases flows. However, we do observe a neat dispersion in the values measured across counties in the same city (Fig.~\ref{fig:transfer}~(g-i)).  
Even so, while in Atlanta we see that each region more or less responded independently in terms of reducing transmission, for the case of Chicago we begin to see a pattern emerging, that becomes clear when looking at NYC where the various regions were clustered temporally and reduced transmission in a monotonically decreasing fashion with mobility-reduction. These type of diagrams allows us to obtain a Pearson correlation $R_{p,response}$ from the scattered plots to account for the synchronization between the counties of an urban area. For example, a city like NYC in Fig.~\ref{fig:transfer}(i) has a larger $R_{p,response}$ than Atlanta in Fig.~\ref{fig:transfer}(g) for which $R_{p,response}$ is close to zero.

\subsection*{ Generalization to a set of cities }

These patterns generalize beyond the three cities considered. In SI Fig. S4 we plot the equivalent of Fig.~\ref{fig:transfer}~(g-i) for the full set of cities, and in Fig.~\ref{fig:cities}~(a) we
plot  $R_{early}$ measured at city scale in the early stages of the pandemic (three
weeks after onset of 100 cases and before the lockdown) as a function of the original $\Phi$.
There is a clear connection between transmissibility in terms of $R_{early}$ and the mobility hierarchy, with hierarchical cities showing an increased spread of the disease at the onset. Indeed NYC being the most hierarchical city in the United States, had a ~50\% higher value of $R_{early}$ than sprawled ones such as Cincinnati, Charlotte or Atlanta. 

This increased transmissibility is also reflected in the extent of the spread of the disease as measured by plotting $I_{\textrm{max}}$ versus the original $\Phi$ in Fig.~\ref{fig:cities}~(b). Cities can be separated into those that had already experienced a peak in the incidence curve (Northeastern and Midwestern cities, colored in yellow), and those that were still at the early phases of the pandemic (Southern and Western cities, in red). Restricting to cities where the pandemic was already well-established, we see a clear trend whereby hierarchical cities had a much wider outbreak as compared to sprawled ones. Stronger mitigation strategies manifested in those places where the outbreak was wide-spread~\cite{colizza2020}, and a relation between mobility reduction and $\Phi$ emerges as can be seen in Fig.~\ref{fig:cities}~(c) where we found that NYC, San Francisco and Boston reduced mobility by around $~60\%$, whereas on the other end of the spectrum, Cincinnati, Minneapolis, Atlanta or Charlotte decreased mobility by around $~40\%$. 

The connection between mobility reduction and transmission mitigation is also far more pronounced: plotting $R_{p,response}$ as a function of $\Phi$ indicates that hierarchical cities seem significantly more responsive to mitigation measures (Fig.~\ref{fig:cities}~(d)). That is, hierarchical cities exhibit more spatially synchronized slowdowns of the transmission after mobility restrictions are applied, while sprawled cities tended to asynchronous evolution at the county level.

\subsection*{Model confirmation of the empirical results}

The empirical results (Figs.~\ref{fig:transfer} and \ref{fig:cities}) suggest that hierarchical cities experience faster and more widespread outbreaks as compared to sprawled ones. However, mobility restrictions and lockdowns in those cities were comparatively more effective. To confirm these findings while removing potential confounding factors as differences in population sizes, densities and spatial distributions, variation in type and timing of lockdowns, or noise in the data, we implement a metapopulation SEIR model driven by empirical mobility flows before and after mitigation measures. 

\begin{figure}
\begin{center}
\includegraphics[width=12cm]{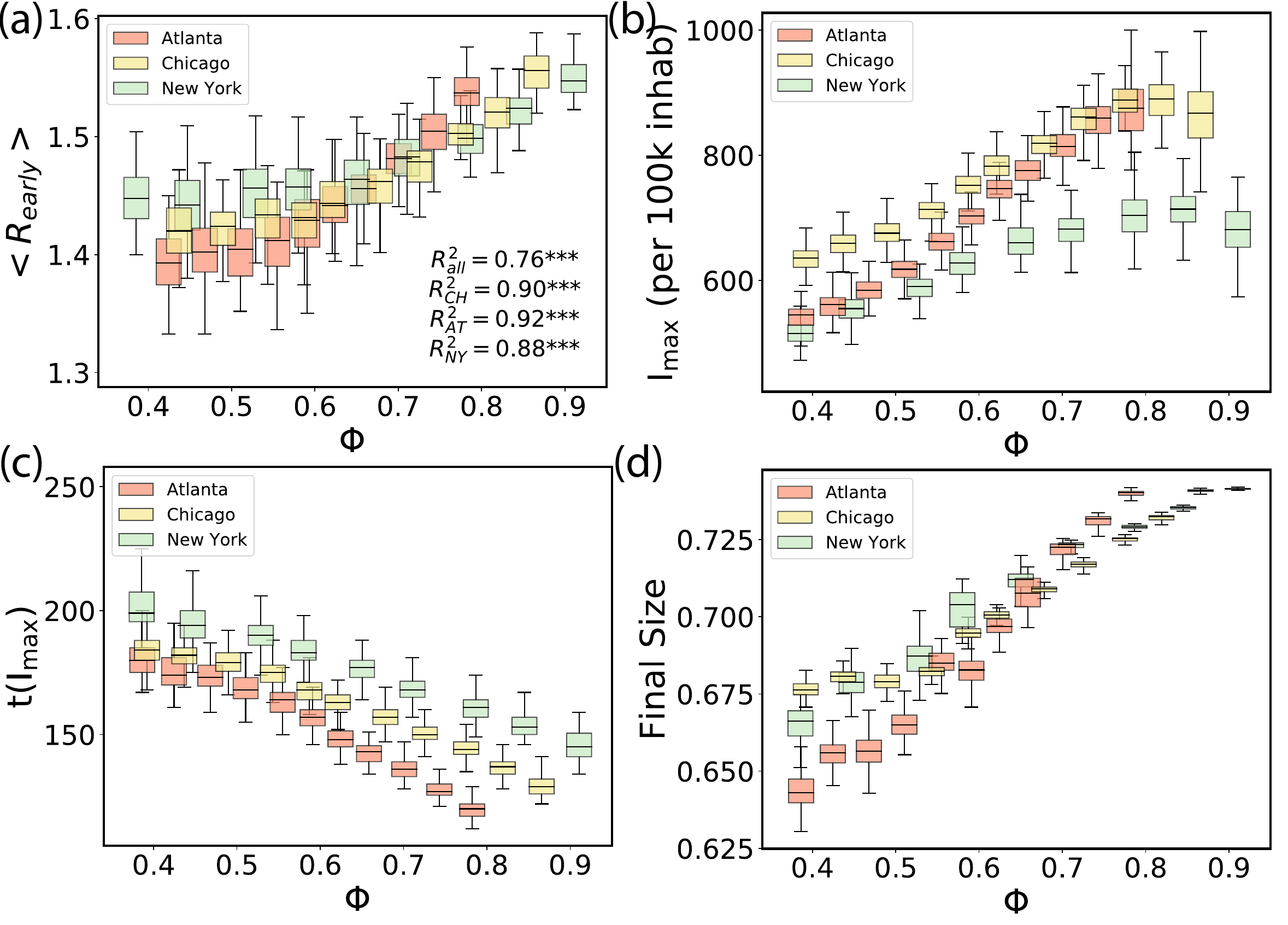}
\caption{Spreading by type of city.  Simulations are run in each city with
  different values of $\Phi$, obtained by the randomization procedure described in the text. Each box reflects 100 runs and displays the median, quartiles, the 5\% and 95\% confidence intervals. In (a), $R_{early}$  (obtained over three weeks after the onset of $10^4$ cases) as a function of $\Phi$, as in Fig. \ref{fig:cities}~(a). The peak incidence $I_{\textrm{max}}$ is shown in  (b), the time to the peak since the beginning of the simulation, $t(I_{max})$, in (c)  and the final epidemic size in (d), all as a function of $\Phi$. All four panels  correspond to the baseline mobility before lockdown. 
  \label{fig:model2}}
\end{center}
\end{figure}

For this, we check if simulating disease spread on multiple idealizations of a city while tuning their mobility structure confirms the empirical observations of the pandemic trends. As representative examples, we consider Atlanta, Chicago and New York and simulate spread while changing $\Phi$ by reshuffling the trip destinations. We end up with thirty cities, ten versions for each of the three, each with a different level of hierarchy score. We begin by checking whether the model can reproduce the trend in Fig.~\ref{fig:cities}~(a) in the early stages of the epidemic without lockdowns. The equivalent plot is shown in Fig.~\ref{fig:model2}~(a), where we extract  $R_{early}$ from the incidence curves of the simulations. In contrast to the empirical data where a significant fraction of cases is undetected, we have complete information in our simulations. Consequently, we set the epidemic onset at $10^4$ accumulated cases. The Pearson coefficient $R^2$ for the points of each individual city ranges between $0.95-0.97$, indicating a clear signal for the transmissibility increasing with hierarchy. When measuring the correlation for the three cities collectively, confounding factors such as differences in population density and mobility distributions emerge, and  $R^2$ decreases to $0.63$ in order of magnitude agreement with the one empirically observed in Fig. \ref{fig:cities}~(a) obtained by aggregating 22 different cities. Similar to Fig.~\ref{fig:cities}~(b), we reproduce the observed trend of the incidence peak $I_{\textrm{max}}$ being monotonic with respect to $\Phi$ as shown in Fig.~\ref{fig:model2}~(b). This is reflected in the time taken to reach the incidence peak (Fig.~\ref{fig:model2}~(c)) and the final size of the outbreak (Fig.~\ref{fig:model2}~(d)), with more centralized-hierarchical cities experiencing stronger transmission, faster spread and wider prevalence in the population. 
 
\subsection*{Modeling the response of cities to mobility restrictions}

\begin{figure}
\begin{center}
\includegraphics[width=12cm]{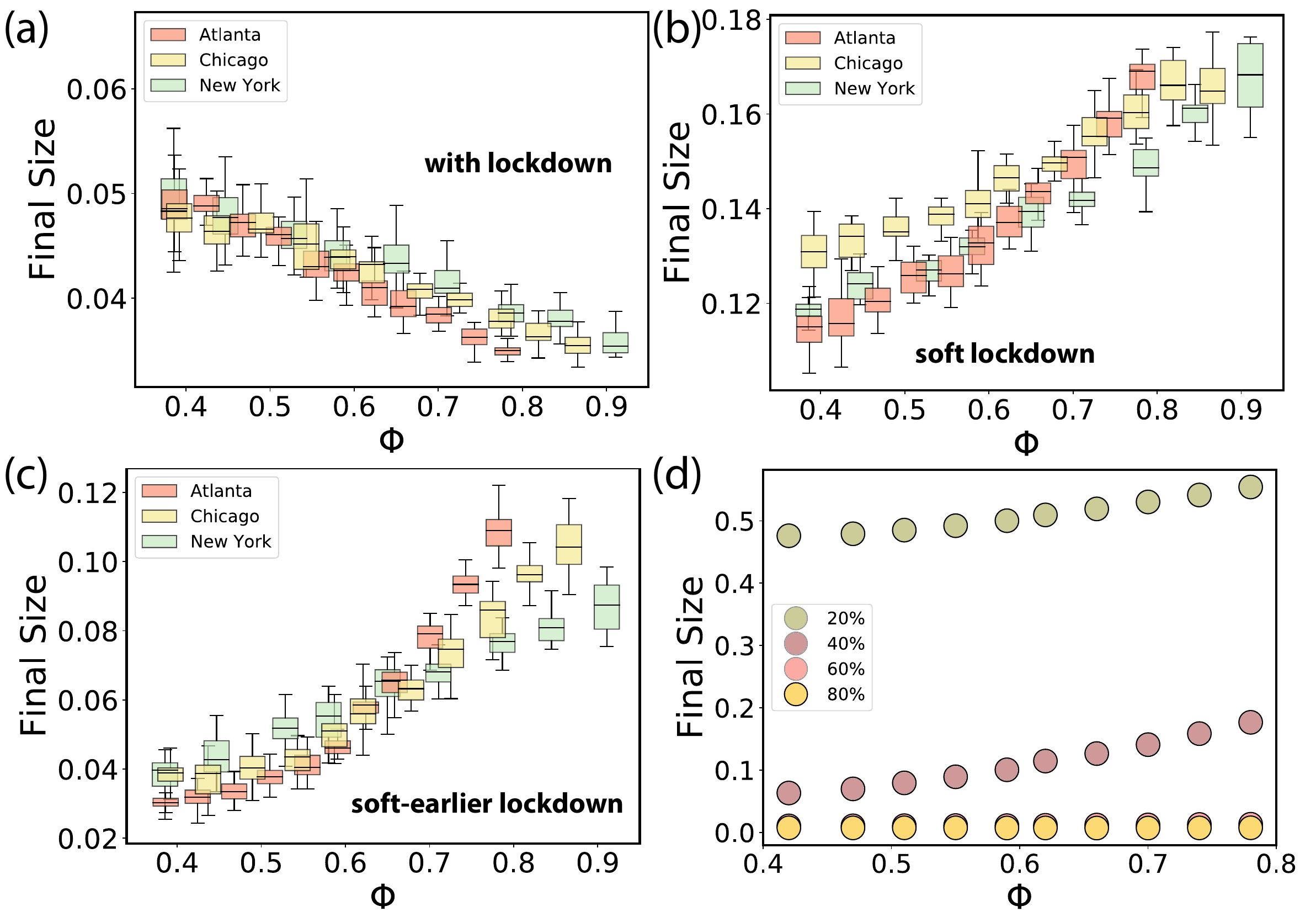}
\caption{Modeling response of cities to mitigation measures. Results
  of a metapopulation model using the S2 cells as basic
  geographical units. Simulations are run in each city with
  different values of $\Phi$, obtained by the randomization procedure described in the text. Each box reflects 100 runs and displays the median, quartiles, the 5\% and 95\% confidence intervals. We explore the pandemic size for different lockdown scenarios: a strong lockdown, $X_S = 0.8$, for $\pi_{\textrm{th}} = 5\times  10^{-3}$ in (a); a soft lockdown, $X_S = 0.4$, at the same prevalence in (b); another soft but earlier lockdown $X_S = 0.4$ and $\pi_{\textrm{th}} = 10^{-3}$ in (c); and finally, in (d), a systematic exploration with $X_S$ for $\pi_{\textrm{th}}= 5\times  10^{-3}$ in Atlanta. 
  \label{fig:model3}}
\end{center}
\end{figure}

On the other hand, Fig.~\ref{fig:cities}~(d) suggests that cities with higher $\Phi$ largely achieved a better reduction in transmission with lockdown measures. To simulate this, in Fig.~\ref{fig:model3}~(a) we plot the final size of the pandemic by establishing an $80$\% reduction in interactions ($X_S = 0.8$) with the lockdown coming into effect at $\pi_\textrm{th} = 5 \times 10^{-3}$. Remarkably, we see an inversion of the curve as compared to Fig.~\ref{fig:model2}~(d) with the trends now reversed; the final size of the pandemic is \textit{lower} in cities with higher $\Phi$ and is decreased by an order of magnitude as compared to the scenario with no mitigation. The inversion of the curve resulted from a rather strong lockdown, so in Fig.~\ref{fig:model3}~(b) we show the case for a softer lockdown with $X_S = 0.4$; here we see the same trend with $\Phi$ as in the case with no lockdown, however, the size of the outbreak is reduced by a factor of five. In addition, the time taken to institute mobility restriction measures is a crucial parameter, and in Fig.~\ref{fig:model3}~(c), we show the case for an earlier (but still soft) lockdown with $X_S = 0.4$ and $\pi_{\textrm{th}} = 10^{-3}$, finding a further decrease by a factor of three in the pandemic size. In terms of assessing the effect of different flavors of lockdowns, the connection between the size of the outbreak and its dependence on $\Phi$, is in general a complex function of the epidemiological parameters, the extent of mobility reduction and the distribution of flows. One can get an impression of this connection by plotting in Fig. ~\ref{fig:model3}~(d) the final epidemic size as a function of $\Phi$ for different lockdown intensities $X_S$. While the curves for $X_S \le 0.4$ do not show an inversion but a progressive flattening, lockdowns with $X_S$ of $0.6$ and above produce the inversion. Although, this occurs for a set of parameters, for a given infectivity $\beta$ and lockdown threshold $\pi_{\textrm{th}}$. 

\subsection*{Robustness of the results}

To confirm the qualitative trends presented thus far, we systematically explore the effects of different parameters on the model results, as well as run our model on a different dataset. A key parameter is the infectivity $\beta$ and, consequently $R_0$.  In SI Figs. S8-S10, the epidemic size is studied as a function of $\Phi$ for different lockdowns $X_S$ while changing $\beta$. Strong lockdowns flatten the curves with $\Phi$ or even invert it for all the considered cases. The same occurs if we vary the times at which lockdowns are imposed through modifying the parameter $\pi_{\textrm{th}}$ as shown in SI Figs. S11 and S12.  Additionally, the results are also robust to the fraction of people leaving geographical units $M$ as shown in SI Figs. S13 and S14. 

The mobility data considered here has information on both commuting and non-commuting flows, all of which can serve as potential infection routes. However, one can repeat the analysis using different types of mobility data, and here we do so by using the publicly available commuting data from the United States census bureau's LODES database for the year 2016 collected at the zip code level. In this case, the number of residents, commuting flows and fractions of mobile population can be directly measured. In SI Figs. S15-S18, we use Atlanta as a representative example and plot $R_{eff}$ at early onset, the maximum incidence $I_{max}$, and the size of the epidemic (for two different flavors of lockdown) as a function of $\Phi$ (recomputed at the zip code resolution and only from commuting data) reproducing the trends seen in Fig.~\ref{fig:model3}.
  
Thus far, we presented a systematic analysis of the parametric landscape of a minimal model that is valid in characterizing the relation between mobility hierarchy and epidemic spreading in cities. In order to focus on the characteristics unique to SARS-CoV-2 we next test the robustness of our findings with a realistic model \cite{domenico2020} that adds age diversification of agents, age-dependent contact patterns for the individuals (POLYMOD matrices), an expanded set of health state compartments and a set of parameters fit by the authors of~\cite{domenico2020} to mimic the real data of daily occupancy in the hospitals of Paris. We employ the same rewiring procedure to scan over multiple values of  $\Phi$. The model is then run with data from Paris and from London using both commuting and census data (SI Figs. S19-S23) as well as the Location History data (Supplementary Figs. S24-S26). In all cases, we observe an increase of $R_{early}$ with $\Phi$, as well as an inversion or flattening of the size vs $\Phi$ lockdown measures, consistent with the results from the minimal model. 

We note that while in our simulations we considered a homogeneous confinement ($X_S$ equal in all S2 cells or zip codes), the mobility reduction as measured directly from data is heterogeneous. Consequently we modify this by assuming a linear relation between mobility reduction by zones and the confinement $X_S$ (SI Figs. S27 and S28). Furthermore, we update the values of $X_S$ weekly according to data, therefore generating restriction measures that are heterogeneous in both space and time. This refinement contains signatures of socioeconomic diversity of neighborhoods, since residents may work in sectors that were impacted differently by the mitigation measures. 
All these tests corroborated our main findings: the effective infectiousness of the COVID-19 depends on the urban substrate, in particular, hierarchical cities will have a more explosive epidemic spreading. The appropriate lockdown measures can flatten the size dependence on $\Phi$ or even invert it (SI Figs. S29 and S30).  

Beyond city scale, at the level of counties, the models are also able to reproduce the dependence of the synchronization of the epidemic curve on the hierarchy of cities (Figs. 1\textbf{g-i} and 2\textbf{d}). In SI Fig. S31, we provide the results aggregated at the county scale in New York. As can be observed, while mobility restrictions are kept the same among all NY counties, the way in which hotspots are connected is critical to the synchronization of local outbreaks and mobility. Reproducing the empirical features requires thus hierarchy in the mobility and heterogeneity in the mobility and in the mitigation measures (confinement). 

In terms of the robustness of empirical analysis, we show in SI Figs. S31-S37 the variation with temporal period of measurement of the epidemic parameters. One could decide to let more delay in the process of checking the effect of mobility restrictions on $R_{eff}$. In Fig. \ref{fig:cities}\textbf{d}, we used a one week delay, in SI Fig. S32 we use a two week delay, finding very similar results. It is true that the value of $R_{eff}$ in early stages can be easily overestimated \cite{starnini2020}, hence in Supplementary Fig. S33 we choose to measure the $R_{early}$ as the average value measured in a time period shifted to one week later with respect to Figure \ref{fig:cities}\textbf{a}. The results remain unaffected. 

The epidemiological data used so far does not report cases by onset of symptoms, but by reporting date. This could introduce some differences \cite{ISSitaly}. We were not able to find a comprehensive data for the US at county scale with onset of symptoms, hence we rely on the UK dataset (https://coronavirus.data.gov.uk/details/download) which reports cases curves for 210 Upper Tier Local Authorities (UTLA), both for reporting and specimen date. What we see in SI Fig. S34 is that the daily estimation of $R_{eff}$ between cases by specimen date and those by publication date tend to differ slightly. However, in our main analysis we do not compute the $R_{eff}$ day by day, but we take for each weak an average value of $R_{eff}$ for all the seven days. This allows us to keep the $R_{eff}$ estimation in a reasonable range lower then $R_{eff}=3$ both for specimen date and publication date cases. This difference among the two estimations of $R_{eff}$ tend to vanish as the weeks go by, showing a higher and higher correlation among all the 210 UTLA. Moreover, in Fig. \ref{fig:cities}  we consider the average of $R_{eff}$ over the first three weeks of cases to generate $R_{early}$  and this leads to SI Fig. S35, where the fluctuations of the first week are mitigated by the subsequent measurements, leading to a much better estimation, which is highly correlated to the one measured on the specimen date. The same issue regarding the reported cases could affect the measure of the incidence peaks in the various cities. Here, by using the same datasets available in UK we use the same definition of incidence peak of Figure \ref{fig:cities}. As shown in SI Fig. S36, the sum of the cases of the week before the peak in the smoothed epidemic curve is pretty consistent among the two definition of cases reporting. Transparency and colormaps are applied proportionally to the local population of the relative UTLA. As can be seen, there are some small deviations among those areas that register low population, in these cases even below $100,000$ inhabitants, hence small fluctuations in the data of low populated areas may strongly affect the measure of the incidence peak with respect to other areas, such as in our case of metropolitan areas. 

Finally, different lockdowns experienced in cities with diverse hierarchies may depend even on socio-economic variables \cite{fraiberger2020}, such as the amount of work force employed in essential services. In Supplementary Fig. 37 we provide a multivariate analysis with socio-economic factors available at the scale of metropolitan areas (OECD defined areas, \url{http://www.oecd.org/cfe/regionaldevelopment/stat.htm}). We do not have a specific variable for the fraction of service sector employed workers, however we have the GDP per capita, which seems to represent a good and reliable proxy for this socio-economic variable (see \url{https://ourworldindata.org/grapher/gdp-vs-services-employment}). As shown in SI Fig. S37, the GDP per capita only explains the $28\%$ of the variance in the distribution of the maximum decrease of mobility in cities, versus the $38\%$ explained by the hierarchy ($\Phi$). For sake of completeness, we provide a further analysis on the effect of hierarchy on the heterogeneity of mobility reductions in city counties. As we can see from SI Fig. S38, more centralized-hierarchical cities tend to experience more homogeneous mobility reductions among their counties during the lockdown, whereas sprawled cities exhibit higher heterogeneity among mobility reductions, causing a harder containment of the local epidemic. The only outlier is Seattle, which presents a peculiar elongated urban shape, where a mobility reduction in a central county necessarily affects all the others.

\section*{Discussion}

In summary, we studied how mobility restrictions in urban areas affect the propagation of an infectious disease. We leveraged a dataset that captures aggregate flows of populations around the world, in a consistent way from the period before the pandemic to a few months after the first wave. In a previous work, we had
shown that hierarchical cities have better indicators in terms of
the use of public transportation, walking, emissions per-capita and
health indicators. However, their mobility structure favors
spreading of infectious diseases in terms of speed and extent of the contagions. At the same time, the empirical analysis shows that lockdown and travel restriction measures can lead to better outcomes in more hierarchical cities. 

Besides empirical observations, we corroborate our findings with a family of metapopulation epidemic models operating at sub-urban scales. The results of the main manuscript are obtained with the simplest version of the models, in which we include the distribution of the population and the mobility information only.  This is enough to be able to generate a dependence between the speed and virulence in terms of maximum of incidence and total disease size, and the mobility hierarchy in cities. Such relation originates from the mobility configuration regarding hotspots, where the most diverse contacts concentrate and thanks to which the contagions can spread faster across the city. Hierarchical cities are characterized by intense and better connected hotspots with respect to decentralized ones. Mitigation measures in the form of mobility reductions impact mainly the hotspot and this is why hierarchical cities respond better to lockdowns. This is the main order effect, other factors such as age structure or socio-economic differences across neighborhoods can play a role as well as observed at coarser scales \cite{bassolas2021diffusion,gauvin2021socio,heroy2021covid,valdano2021}. We include these effects in more elaborate versions of the model and find in all the cases that the relation between hierarchy and spreading speed and the possibility of a better response of hierarchical cities to lockdowns still hold, which means that those factors induce changes of higher order in the propagation patterns. These results have been confirmed using a set of US cities as basis for the simulations, their randomized clones, and also data from London and Paris. Still the relation between mobility hierarchy and spreading is robust and holds too in other cities and even in theoretical settings such as lattices. 

Several implications present themselves from a policy making point of view: Given the limited resources for testing and surveillance in much of the developing world (and for that matter in parts of the developed world as well), it seems prudent to classify cities that are more vulnerable to spread ahead of time and deploy those resources there first. Furthermore, the analysis provides clues on the extent to which non-pharmaceutical interventions are effective. It is effective to enforce mitigation
measures as early and as thoroughly as possible, given that the
time-to-response is particularly  crucial in hierarchical cities within or near an epidemic outbreak. Sprawled cities, with their distributed mobility and less connected outbreaks, have a larger time window within which to enforce policy measures, yet even so, the intensity of response is important to reduce the final number of cases. Similar lessons may be employed in terms of vaccination strategies, given its even more limited availability. While well established graduated procedures for vaccinations have been followed worldwide, in the face of exponential growth as has being seen in waves in India and Brazil, it may be more effective to deploy vaccines considering mobility hotspots (e.g., service workers), motivating new research in optimal spatial vaccination strategies. We note that our findings, while presented in the context of COVID-19, are generally applicable to other respiratory infectious diseases. In terms of a post-pandemic urban policies, just as outbreaks of tuberculosis and other diseases shaped the form of the modern urban space in the past, our results also shed new light on how to protect the different type of cities from infectious disease spreading.

\section*{Acknowledgements}
We thank Aaron Schneider, Aaron Stein, Ahmed Aktay, Alvin Raj, Amy Chung-Yu Chou, Andrew Oplinger, Ashley Zlatinov, Blaise Aguera y Arcas, Bryant Gipson, Charina Chou, Christopher Pluntke, Damien Desfontaines, Eric Tholome, Ewa Dominowska, Gregor Rothfuss, Iz Conroy, Janel Thamkul, Janet Whiteman, Jason Freidenfelds, Jeff Dean, Karen Lee Smith, Katherine Chou, Leeron Morad, Lizzie Dorfman, Marlo McGriff, Mia Vu, Michael Howell, Paul Eastham, Rif Saurous, Rishi Bal, Royce Wilson, Ruth Alcantara, Shawn O'Banion, Stephanie Cason, Thomas Roessler, Vivien Hoang, Yanning Zhang, Xue Ben and Brian Dickinson for their support and guidance.

M.M. is funded by the Conselleria d'Innovaci\'o, Recerca i Turisme of
the Government of the Balearic Islands and the European Social Fund
with grant code FPI/2090/2018. J.A., M.M., S.M. and J.J.R. also
acknowledge funding from the project Distancia-COVID (CSIC-COVID-19)
of the CSIC funded by a contribution of AENA, from the project PACSS RTI2018-093732-B-C22 of the MCIN/AEI/10.13039/501100011033/ and by EU through FEDER funds (A way to make Europe), and also from the Maria de Maeztu program MDM-2017-0711 of the MCIN/AEI/10.13039/501100011033/.
M.M. acknowledges the financial support of the Sorbonne Universit\'e Emergence project RISKFLOW.
A.B. and V.N. acknowledge support from the UK EPSRC New Investigator Award Grant No. EP/S027920/1. G.G., S.H. and S. Mimar acknowledge support from from NSF Grant IIS-2029095 and the US Army Research Office under Agreement Number W911NF-18-1-0421. A.K. is supported by the National Defense Science and Engineering Graduate Fellowship through the Department of Defense.

\section*{Author contributions}

G.G., S. Meloni, V.N.,  J.J.R. and A.S. developed the concepts and designed
the study. A.B., S.H., A.K., M.M. and S. Mimar analyzed the data. A.S.
computed and provided the mobility map data. J.A., S. Meloni and J.J.R.
developed the model. J.A. performed the model simulations. G.G., M.M.,
S. Meloni, J.J.R. and A.S. contributed to the work methodology. G.G., M.M., S.
Meloni, V.N. and J.J.R wrote the paper. G.G., S. Meloni, J.J.R. and A.S. coordinated the
study. All authors read, edited, and approved the final version of the
paper. The authors are listed in alphabetical order.

\section*{Competing Interests}

The authors declare no competing interests.

\section*{Code availability statment}
The code to calculate the flow hierarchy in cities is available in the following link: \url{https://mygit.katolaz.net/aleix/flow-hierarchy}.

\section*{Data availability statement}

\subsection*{Mobility data}

The Google COVID-19 Aggregated Mobility Research Dataset used for this study is available with permission from Google LLC. The data should be interpreted in light of several
limitations. First, the Google mobility data is limited to smartphone
users who have opted in to Googles Location History feature. These data may not be representative of the population as a whole, and furthermore their representativeness may vary by
location. Importantly, these limited data are only viewed through the
lens of differential privacy algorithms, specifically designed to
protect user anonymity and obscure fine detail. 
Moreover, comparisons of direct flows across rather than within locations (particularly different countries) are only descriptive since these regions may potentially differ in substantial ways. 

The commuting data in the US corresponds to the LEHD Origin-Destination Employment Statistics (LODES) dataset of 2016-2017 at zip code level collected by the US census office (\url{https://lehd.ces.census.gov/data/}). The commuting data for London was obtained from \url{https://www.nomisweb.co.uk/census/2011/wu01ew}. In Paris, the commuting information was obtained from \url{https://www.insee.fr/fr/statistiques/4509353} and \url{https://www.insee.fr/fr/statistiques/4509360}.

\subsection*{Demographic data}

Socioeconomic variables in the 22 US cities under study are found in OECD defined urban areas statistics \url{http://www.oecd.org/cfe/regionaldevelopment/stat.htm}. In London, the population by age was obtained from  \url{https://www.ons.gov.uk/peoplepopulationandcommunity/populationandmigration/populationestimates} and in Paris from \url{https://www.insee.fr/fr/statistiques/4515565?sommaire=4516122}.

\subsection*{Epidemic records}

The sources for COVID-19 data are two for the US: The New York Times \url{https://github.com/nytimes/covid-19-data} and USAFacts \url{https://usafacts.org/visualizations/coronavirus-covid-19-spread-map/}, and one for the UK: UTLA \url{https://coronavirus.data.gov.uk/details/download}.

\subsection*{Geographical data}

The city boundaries used throughout the study are the functional urban areas provided by the OECD \url{https://ec.europa.eu/eurostat/statistics-explained/index.php/Glossary:Functional_urban_area}.

\bibliography{mob.bib}

\begin{thebibliography}{10}
\urlstyle{rm}
\expandafter\ifx\csname url\endcsname\relax
  \def\url#1{\texttt{#1}}\fi
\expandafter\ifx\csname urlprefix\endcsname\relax\def\urlprefix{URL }\fi
\expandafter\ifx\csname doiprefix\endcsname\relax\def\doiprefix{DOI: }\fi
\providecommand{\bibinfo}[2]{#2}
\providecommand{\eprint}[2][]{\url{#2}}

\bibitem{Zhou_2020}
\bibinfo{author}{Zhou, P.} \emph{et~al.}
\newblock \bibinfo{journal}{\bibinfo{title}{A pneumonia outbreak associated
  with a new coronavirus of probable bat origin}}.
\newblock {\emph{\JournalTitle{Nature}}} \textbf{\bibinfo{volume}{579}},
  \bibinfo{pages}{270--273}, \doiprefix\url{10.1038/s41586-020-2012-7}
  (\bibinfo{year}{2020}).

\bibitem{Wu_2020}
\bibinfo{author}{Wu, F.} \emph{et~al.}
\newblock \bibinfo{journal}{\bibinfo{title}{A new coronavirus associated with
  human respiratory disease in {China}}}.
\newblock {\emph{\JournalTitle{Nature}}} \textbf{\bibinfo{volume}{579}},
  \bibinfo{pages}{265--269}, \doiprefix\url{10.1038/s41586-020-2008-3}
  (\bibinfo{year}{2020}).

\bibitem{fang2020}
\bibinfo{author}{Fang, H.}, \bibinfo{author}{Wang, L.} \&
  \bibinfo{author}{Yang, Y.}
\newblock \bibinfo{title}{Human mobility restrictions and the spread of the
  novel coronavirus (2019-ncov) in {China}}.
\newblock \bibinfo{type}{Tech. Rep.}, \bibinfo{institution}{National Bureau of
  Economic Research} (\bibinfo{year}{2020}).

\bibitem{haug2020}
\bibinfo{author}{Haug, N.} \emph{et~al.}
\newblock \bibinfo{journal}{\bibinfo{title}{Ranking the effectiveness of
  worldwide {COVID-19} government interventions}}.
\newblock {\emph{\JournalTitle{Nature Human Behaviour}}}
  \textbf{\bibinfo{volume}{4}}, \bibinfo{pages}{1303--1312}
  (\bibinfo{year}{2020}).

\bibitem{pepe2020}
\bibinfo{author}{Pepe, E.} \emph{et~al.}
\newblock \bibinfo{journal}{\bibinfo{title}{Covid-19 outbreak response, a
  dataset to assess mobility changes in italy following national lockdown}}.
\newblock {\emph{\JournalTitle{Scientific data}}} \textbf{\bibinfo{volume}{7}},
  \bibinfo{pages}{1--7} (\bibinfo{year}{2020}).

\bibitem{chinazzi2020}
\bibinfo{author}{Chinazzi, M.} \emph{et~al.}
\newblock \bibinfo{journal}{\bibinfo{title}{The effect of travel restrictions
  on the spread of the 2019 novel coronavirus (covid-19) outbreak}}.
\newblock {\emph{\JournalTitle{Science}}} \textbf{\bibinfo{volume}{368}},
  \bibinfo{pages}{395--400} (\bibinfo{year}{2020}).

\bibitem{dehning2020}
\bibinfo{author}{Dehning, J.} \emph{et~al.}
\newblock \bibinfo{journal}{\bibinfo{title}{Inferring change points in the
  spread of {COVID-19} reveals the effectiveness of interventions}}.
\newblock {\emph{\JournalTitle{Science}}} \textbf{\bibinfo{volume}{369}},
  \bibinfo{pages}{160}, \doiprefix\url{10.1126/science.abb9789}
  (\bibinfo{year}{2020}).

\bibitem{gatto2020}
\bibinfo{author}{Gatto, M.} \emph{et~al.}
\newblock \bibinfo{journal}{\bibinfo{title}{Spread and dynamics of the covid-19
  epidemic in italy: Effects of emergency containment measures}}.
\newblock {\emph{\JournalTitle{Procs. Nat. Acad. Sci. U.S.A.}}}
  \textbf{\bibinfo{volume}{117}}, \bibinfo{pages}{10484--10491},
  \doiprefix\url{10.1073/pnas.2004978117} (\bibinfo{year}{2020}).

\bibitem{kraemer2020}
\bibinfo{author}{Kraemer, M.~U.} \emph{et~al.}
\newblock \bibinfo{journal}{\bibinfo{title}{The effect of human mobility and
  control measures on the {COVID-19 epidemic in China}}}.
\newblock {\emph{\JournalTitle{Science}}} \textbf{\bibinfo{volume}{368}},
  \bibinfo{pages}{493--497} (\bibinfo{year}{2020}).

\bibitem{maier2020}
\bibinfo{author}{Maier, B.~F.} \& \bibinfo{author}{Brockmann, D.}
\newblock \bibinfo{journal}{\bibinfo{title}{Effective containment explains
  subexponential growth in recent confirmed {COVID-19} cases in {China}}}.
\newblock {\emph{\JournalTitle{Science}}} \textbf{\bibinfo{volume}{368}},
  \bibinfo{pages}{742--746} (\bibinfo{year}{2020}).

\bibitem{melo2021}
\bibinfo{author}{Melo, H.~P.} \emph{et~al.}
\newblock \bibinfo{journal}{\bibinfo{title}{Heterogeneous impact of a lockdown
  on inter-municipality mobility}}.
\newblock {\emph{\JournalTitle{Physical Review Research}}}
  \textbf{\bibinfo{volume}{3}}, \bibinfo{pages}{013032} (\bibinfo{year}{2021}).

\bibitem{pan2020}
\bibinfo{author}{Pan, A.} \emph{et~al.}
\newblock \bibinfo{journal}{\bibinfo{title}{Association of public health
  interventions with the epidemiology of the covid-19 outbreak in {Wuhan,
  China}}}.
\newblock {\emph{\JournalTitle{Jama}}}  (\bibinfo{year}{2020}).

\bibitem{schlosser2020}
\bibinfo{author}{Schlosser, F.} \emph{et~al.}
\newblock \bibinfo{journal}{\bibinfo{title}{{COVID-19} lockdown induces
  disease-mitigating structural changes in mobility networks}}.
\newblock {\emph{\JournalTitle{Proceedings of the National Academy of
  Sciences}}} \textbf{\bibinfo{volume}{117}}, \bibinfo{pages}{32883--32890}
  (\bibinfo{year}{2020}).

\bibitem{oh2021mobility}
\bibinfo{author}{Oh, J.} \emph{et~al.}
\newblock \bibinfo{journal}{\bibinfo{title}{Mobility restrictions were
  associated with reductions in covid-19 incidence early in the pandemic:
  evidence from a real-time evaluation in 34 countries}}.
\newblock {\emph{\JournalTitle{Scientific Reports}}}
  \textbf{\bibinfo{volume}{11}}, \bibinfo{pages}{1--17} (\bibinfo{year}{2021}).

\bibitem{flahault1992}
\bibinfo{author}{Flahault, A.} \& \bibinfo{author}{Valleron, A.-J.}
\newblock \bibinfo{journal}{\bibinfo{title}{A method for assessing the global
  spread of hiv-1 infection based on air travel}}.
\newblock {\emph{\JournalTitle{Mathematical Population Studies}}}
  \textbf{\bibinfo{volume}{3}}, \bibinfo{pages}{161--171}
  (\bibinfo{year}{1992}).

\bibitem{grais2003}
\bibinfo{author}{Grais, R.~F.}, \bibinfo{author}{Ellis, J.~H.} \&
  \bibinfo{author}{Glass, G.~E.}
\newblock \bibinfo{journal}{\bibinfo{title}{Assessing the impact of airline
  travel on the geographic spread of pandemic influenza}}.
\newblock {\emph{\JournalTitle{European Journal of Epidemiology}}}
  \textbf{\bibinfo{volume}{18}}, \bibinfo{pages}{1065--1072}
  (\bibinfo{year}{2003}).

\bibitem{hufnagel2004}
\bibinfo{author}{Hufnagel, L.}, \bibinfo{author}{Brockmann, D.} \&
  \bibinfo{author}{Geisel, T.}
\newblock \bibinfo{journal}{\bibinfo{title}{Forecast and control of epidemics
  in a globalized world}}.
\newblock {\emph{\JournalTitle{Procs. Nat. Acad. Sci. U.S.A.}}}
  \textbf{\bibinfo{volume}{101}}, \bibinfo{pages}{15124--15129}
  (\bibinfo{year}{2004}).

\bibitem{brownstein2006}
\bibinfo{author}{Brownstein, J.~S.}, \bibinfo{author}{Wolfe, C.~J.} \&
  \bibinfo{author}{Mandl, K.~D.}
\newblock \bibinfo{journal}{\bibinfo{title}{Empirical evidence for the effect
  of airline travel on inter-regional influenza spread in the united states}}.
\newblock {\emph{\JournalTitle{PLOS Medicine}}} \textbf{\bibinfo{volume}{3}},
  \bibinfo{pages}{e401} (\bibinfo{year}{2006}).

\bibitem{colizza2006}
\bibinfo{author}{Colizza, V.}, \bibinfo{author}{Barrat, A.},
  \bibinfo{author}{Barth{\'e}lemy, M.} \& \bibinfo{author}{Vespignani, A.}
\newblock \bibinfo{journal}{\bibinfo{title}{The role of the airline
  transportation network in the prediction and predictability of global
  epidemics}}.
\newblock {\emph{\JournalTitle{Procs. Nat. Acad. Sci. U.S.A.}}}
  \textbf{\bibinfo{volume}{103}}, \bibinfo{pages}{2015--2020}
  (\bibinfo{year}{2006}).

\bibitem{colizza2007}
\bibinfo{author}{Colizza, V.}, \bibinfo{author}{Barrat, A.},
  \bibinfo{author}{Barthelemy, M.}, \bibinfo{author}{Valleron, A.-J.} \&
  \bibinfo{author}{Vespignani, A.}
\newblock \bibinfo{journal}{\bibinfo{title}{Modeling the worldwide spread of
  pandemic influenza: baseline case and containment interventions}}.
\newblock {\emph{\JournalTitle{PLOS Medicine}}} \textbf{\bibinfo{volume}{4}}
  (\bibinfo{year}{2007}).

\bibitem{tatem2012}
\bibinfo{author}{Tatem, A.} \emph{et~al.}
\newblock \bibinfo{journal}{\bibinfo{title}{Air travel and vector-borne disease
  movement}}.
\newblock {\emph{\JournalTitle{Parasitology}}} \textbf{\bibinfo{volume}{139}},
  \bibinfo{pages}{1816--1830} (\bibinfo{year}{2012}).

\bibitem{finger2016}
\bibinfo{author}{Finger, F.} \emph{et~al.}
\newblock \bibinfo{journal}{\bibinfo{title}{Mobile phone data highlights the
  role of mass gatherings in the spreading of cholera outbreaks}}.
\newblock {\emph{\JournalTitle{Procs. Nat. Acad. Sci. U.S.A.}}}
  \textbf{\bibinfo{volume}{113}}, \bibinfo{pages}{6421--6426}
  (\bibinfo{year}{2016}).

\bibitem{zhang2017}
\bibinfo{author}{Zhang, Q.} \emph{et~al.}
\newblock \bibinfo{journal}{\bibinfo{title}{Spread of zika virus in the
  americas}}.
\newblock {\emph{\JournalTitle{Procs. Nat. Acad. Sci. U.S.A.}}}
  \textbf{\bibinfo{volume}{114}}, \bibinfo{pages}{E4334--E4343}
  (\bibinfo{year}{2017}).

\bibitem{barbosa2018}
\bibinfo{author}{Barbosa, H.} \emph{et~al.}
\newblock \bibinfo{journal}{\bibinfo{title}{Human mobility: Models and
  applications}}.
\newblock {\emph{\JournalTitle{Physics Reports}}}
  \textbf{\bibinfo{volume}{734}}, \bibinfo{pages}{1--74}
  (\bibinfo{year}{2018}).

\bibitem{massaro2019}
\bibinfo{author}{Massaro, E.}, \bibinfo{author}{Kondor, D.} \&
  \bibinfo{author}{Ratti, C.}
\newblock \bibinfo{journal}{\bibinfo{title}{Assessing the interplay between
  human mobility and mosquito borne diseases in urban environments}}.
\newblock {\emph{\JournalTitle{Scientific Reports}}}
  \textbf{\bibinfo{volume}{9}}, \bibinfo{pages}{1--13} (\bibinfo{year}{2019}).

\bibitem{mu2020}
\bibinfo{author}{Mu, X.}, \bibinfo{author}{Yeh, A. G.-O.} \&
  \bibinfo{author}{Zhang, X.}
\newblock \bibinfo{journal}{\bibinfo{title}{{The interplay of spatial spread of
  COVID-19 and human mobility in the urban system of China during the Chinese
  New Year}}}.
\newblock {\emph{\JournalTitle{Environment and Planning B: Urban Analytics and
  City Science}}} \bibinfo{pages}{2399808320954211} (\bibinfo{year}{2020}).

\bibitem{niu2020}
\bibinfo{author}{Niu, X.}, \bibinfo{author}{Yue, Y.}, \bibinfo{author}{Zhou,
  X.} \& \bibinfo{author}{Zhang, X.}
\newblock \bibinfo{journal}{\bibinfo{title}{How urban factors affect the
  spatiotemporal distribution of infectious diseases in addition to intercity
  population movement in {China}}}.
\newblock {\emph{\JournalTitle{ISPRS International Journal of
  Geo-Information}}} \textbf{\bibinfo{volume}{9}}, \bibinfo{pages}{615}
  (\bibinfo{year}{2020}).

\bibitem{balcan2009}
\bibinfo{author}{Balcan, D.} \emph{et~al.}
\newblock \bibinfo{journal}{\bibinfo{title}{Multiscale mobility networks and
  the spatial spreading of infectious diseases}}.
\newblock {\emph{\JournalTitle{{Procs. Natl. Acad. Sci. U.S.A.}}}}
  \textbf{\bibinfo{volume}{106}}, \bibinfo{pages}{21484--21489},
  \doiprefix\url{10.1073/pnas.0906910106} (\bibinfo{year}{2009}).

\bibitem{gilbert2020}
\bibinfo{author}{Gilbert, M.} \emph{et~al.}
\newblock \bibinfo{journal}{\bibinfo{title}{Preparedness and vulnerability of
  african countries against importations of covid-19: a modelling study}}.
\newblock {\emph{\JournalTitle{The Lancet}}} \textbf{\bibinfo{volume}{395}},
  \bibinfo{pages}{871--877} (\bibinfo{year}{2020}).

\bibitem{rader2020}
\bibinfo{author}{Rader, B.} \emph{et~al.}
\newblock \bibinfo{journal}{\bibinfo{title}{Crowding and the shape of
  {COVID-19} epidemics}}.
\newblock {\emph{\JournalTitle{Nature medicine}}}
  \textbf{\bibinfo{volume}{26}}, \bibinfo{pages}{1829--1834}
  (\bibinfo{year}{2020}).

\bibitem{mazzoli2021interplay}
\bibinfo{author}{Mazzoli, M.} \emph{et~al.}
\newblock \bibinfo{journal}{\bibinfo{title}{Interplay between mobility,
  multi-seeding and lockdowns shapes covid-19 local impact}}.
\newblock {\emph{\JournalTitle{PLoS Computational Biology}}}
  \textbf{\bibinfo{volume}{17}}, \bibinfo{pages}{e1009326}
  (\bibinfo{year}{2021}).

\bibitem{kraemer2021spatiotemporal}
\bibinfo{author}{Kraemer, M.~U.} \emph{et~al.}
\newblock \bibinfo{journal}{\bibinfo{title}{Spatiotemporal invasion dynamics of
  sars-cov-2 lineage b. 1.1. 7 emergence}}.
\newblock {\emph{\JournalTitle{Science}}} \textbf{\bibinfo{volume}{373}},
  \bibinfo{pages}{889--895} (\bibinfo{year}{2021}).

\bibitem{ferguson2005}
\bibinfo{author}{Ferguson, N.~M.} \emph{et~al.}
\newblock \bibinfo{journal}{\bibinfo{title}{Strategies for containing an
  emerging influenza pandemic in southeast asia}}.
\newblock {\emph{\JournalTitle{Nature}}} \textbf{\bibinfo{volume}{437}},
  \bibinfo{pages}{209--214} (\bibinfo{year}{2005}).

\bibitem{hollingsworth2006}
\bibinfo{author}{Hollingsworth, T.~D.}, \bibinfo{author}{Ferguson, N.~M.} \&
  \bibinfo{author}{Anderson, R.~M.}
\newblock \bibinfo{journal}{\bibinfo{title}{Will travel restrictions control
  the international spread of pandemic influenza?}}
\newblock {\emph{\JournalTitle{Nature Medicine}}}
  \textbf{\bibinfo{volume}{12}}, \bibinfo{pages}{497--499}
  (\bibinfo{year}{2006}).

\bibitem{epstein2007}
\bibinfo{author}{Epstein, J.~M.} \emph{et~al.}
\newblock \bibinfo{journal}{\bibinfo{title}{Controlling pandemic flu: the value
  of international air travel restrictions}}.
\newblock {\emph{\JournalTitle{PLOS ONE}}} \textbf{\bibinfo{volume}{2}},
  \bibinfo{pages}{e401} (\bibinfo{year}{2007}).

\bibitem{bajardi2011}
\bibinfo{author}{Bajardi, P.} \emph{et~al.}
\newblock \bibinfo{journal}{\bibinfo{title}{Human mobility networks, travel
  restrictions, and the global spread of 2009 {H1N1} pandemic}}.
\newblock {\emph{\JournalTitle{PLOS ONE}}} \textbf{\bibinfo{volume}{6}},
  \bibinfo{pages}{e16591}, \doiprefix\url{10.1371/journal.pone.0016591}
  (\bibinfo{year}{2011}).

\bibitem{meloni2011}
\bibinfo{author}{Meloni, S.} \emph{et~al.}
\newblock \bibinfo{journal}{\bibinfo{title}{Modeling human mobility responses
  to the large-scale spreading of infectious diseases}}.
\newblock {\emph{\JournalTitle{Scientific Reports}}}
  \textbf{\bibinfo{volume}{1}}, \bibinfo{pages}{62} (\bibinfo{year}{2011}).

\bibitem{poletto2014}
\bibinfo{author}{Poletto, C.} \emph{et~al.}
\newblock \bibinfo{journal}{\bibinfo{title}{Assessing the impact of travel
  restrictions on international spread of the 2014 west african ebola
  epidemic}}.
\newblock {\emph{\JournalTitle{Eurosurveillance}}}
  \textbf{\bibinfo{volume}{19}}, \bibinfo{pages}{20936} (\bibinfo{year}{2014}).

\bibitem{pei2020}
\bibinfo{author}{Pei, S.}, \bibinfo{author}{Kandula, S.} \&
  \bibinfo{author}{Shaman, J.}
\newblock \bibinfo{journal}{\bibinfo{title}{Differential effects of
  intervention timing on covid-19 spread in the united states}}.
\newblock {\emph{\JournalTitle{Science Advances}}}
  \textbf{\bibinfo{volume}{6}}, \bibinfo{pages}{eabd6370}
  (\bibinfo{year}{2020}).

\bibitem{varga2016}
\bibinfo{author}{Varga, L.}, \bibinfo{author}{Kov\'acs, A.},
  \bibinfo{author}{T\'oth, G.}, \bibinfo{author}{Papp, I.} \&
  \bibinfo{author}{N\'eda, Z.}
\newblock \bibinfo{journal}{\bibinfo{title}{Further we travel the faster we
  go}}.
\newblock {\emph{\JournalTitle{PLOS ONE}}} \textbf{\bibinfo{volume}{11}},
  \bibinfo{pages}{e0148913}, \doiprefix\url{10.1371/journal.pone.0148913}
  (\bibinfo{year}{2016}).

\bibitem{Lee_2017}
\bibinfo{author}{Lee, M.}, \bibinfo{author}{Barbosa, H.},
  \bibinfo{author}{Youn, H.}, \bibinfo{author}{Holme, P.} \&
  \bibinfo{author}{Ghoshal, G.}
\newblock \bibinfo{journal}{\bibinfo{title}{Morphology of travel routes and the
  organization of cities}}.
\newblock {\emph{\JournalTitle{{Nature Communications}}}}
  \textbf{\bibinfo{volume}{8}}, \bibinfo{pages}{1--10} (\bibinfo{year}{2017}).

\bibitem{Kirkley_2018}
\bibinfo{author}{Kirkley, A.}, \bibinfo{author}{Barbosa, H.},
  \bibinfo{author}{Barthelemy, M.} \& \bibinfo{author}{Ghoshal, G.}
\newblock \bibinfo{journal}{\bibinfo{title}{From the betweenness centrality in
  street networks to structural invariants in random planar graphs}}.
\newblock {\emph{\JournalTitle{Nature Communications}}}
  \textbf{\bibinfo{volume}{9}}, \bibinfo{pages}{2501} (\bibinfo{year}{2018}).

\bibitem{soriano2018}
\bibinfo{author}{Soriano-Pa\~nos, D.}, \bibinfo{author}{Lotero, L.},
  \bibinfo{author}{Arenas, A.} \& \bibinfo{author}{G\'omez-Garde\~nes, J.}
\newblock \bibinfo{journal}{\bibinfo{title}{Spreading processes in multiplex
  metapopulations containing different mobility networks}}.
\newblock {\emph{\JournalTitle{Phys. Rev. X}}} \textbf{\bibinfo{volume}{8}},
  \bibinfo{pages}{031039}, \doiprefix\url{10.1103/PhysRevX.8.031039}
  (\bibinfo{year}{2018}).

\bibitem{bassolas2021diffusion}
\bibinfo{author}{Bassolas, A.}, \bibinfo{author}{Sousa, S.} \&
  \bibinfo{author}{Nicosia, V.}
\newblock \bibinfo{journal}{\bibinfo{title}{Diffusion segregation and the
  disproportionate incidence of covid-19 in african american communities}}.
\newblock {\emph{\JournalTitle{Journal of the Royal Society Interface}}}
  \textbf{\bibinfo{volume}{18}}, \bibinfo{pages}{20200961}
  (\bibinfo{year}{2021}).

\bibitem{gauvin2021socio}
\bibinfo{author}{Gauvin, L.} \emph{et~al.}
\newblock \bibinfo{journal}{\bibinfo{title}{Socioeconomic determinants of
  mobility responses during the first wave of {COVID-19} in {Italy}: from
  provinces to neighbourhoods}}.
\newblock {\emph{\JournalTitle{Journal of the Royal Society Interface}}}
  \textbf{\bibinfo{volume}{18}}, \bibinfo{pages}{20210092}
  (\bibinfo{year}{2021}).

\bibitem{heroy2021covid}
\bibinfo{author}{Heroy, S.}, \bibinfo{author}{Loaiza, I.},
  \bibinfo{author}{Pentland, A.} \& \bibinfo{author}{O’Clery, N.}
\newblock \bibinfo{journal}{\bibinfo{title}{Covid-19 policy analysis: labour
  structure dictates lockdown mobility behaviour}}.
\newblock {\emph{\JournalTitle{Journal of the Royal Society Interface}}}
  \textbf{\bibinfo{volume}{18}}, \bibinfo{pages}{20201035}
  (\bibinfo{year}{2021}).

\bibitem{valdano2021}
\bibinfo{author}{Valdano, E.}, \bibinfo{author}{Lee, J.},
  \bibinfo{author}{Bansal, S.}, \bibinfo{author}{Rubrichi, S.} \&
  \bibinfo{author}{Colizza, V.}
\newblock \bibinfo{journal}{\bibinfo{title}{Highlighting socio-economic
  constraints on mobility reductions during covid-19 restrictions in france can
  inform effective and equitable pandemic response}}.
\newblock {\emph{\JournalTitle{Journal of travel medicine}}}
  \textbf{\bibinfo{volume}{28}}, \bibinfo{pages}{taab045}
  (\bibinfo{year}{2021}).

\bibitem{banai2020}
\bibinfo{author}{Banai, R.}
\newblock \bibinfo{journal}{\bibinfo{title}{Pandemic and the planning of
  resilient cities and regions}}.
\newblock {\emph{\JournalTitle{Cities}}} \textbf{\bibinfo{volume}{106}},
  \bibinfo{pages}{102929} (\bibinfo{year}{2020}).

\bibitem{de2000fear}
\bibinfo{author}{De~La~Barra, X.}
\newblock \bibinfo{journal}{\bibinfo{title}{Fear of epidemics: the engine of
  urban planning}}.
\newblock {\emph{\JournalTitle{Planning practice \& research}}}
  \textbf{\bibinfo{volume}{15}}, \bibinfo{pages}{7--16} (\bibinfo{year}{2000}).

\bibitem{eltarabily2020post}
\bibinfo{author}{Eltarabily, S.} \& \bibinfo{author}{Elghezanwy, D.}
\newblock \bibinfo{journal}{\bibinfo{title}{Post-pandemic cities-the impact of
  covid-19 on cities and urban design}}.
\newblock {\emph{\JournalTitle{Architecture Research}}}
  \textbf{\bibinfo{volume}{10}}, \bibinfo{pages}{75--84}
  (\bibinfo{year}{2020}).

\bibitem{martinez2021}
\bibinfo{author}{Mart{\'\i}nez, L.} \& \bibinfo{author}{Short, J.~R.}
\newblock \bibinfo{journal}{\bibinfo{title}{The pandemic city: Urban issues in
  the time of covid-19}}.
\newblock {\emph{\JournalTitle{Sustainability}}} \textbf{\bibinfo{volume}{13}},
  \bibinfo{pages}{3295} (\bibinfo{year}{2021}).

\bibitem{unurban}
\bibinfo{author}{Report:, U.~N.}
\newblock \bibinfo{title}{$68\%$ of the world population projected to live in
  urban areas by 2050} (\bibinfo{year}{2018}).

\bibitem{grant2020cities}
\bibinfo{author}{Grant, J.}
\newblock \bibinfo{journal}{\bibinfo{title}{What cities can learn from lockdown
  about planning for life after the coronavirus pandemic}}.
\newblock {\emph{\JournalTitle{The Conversation}}}  (\bibinfo{year}{2020}).

\bibitem{dalziel2018}
\bibinfo{author}{Dalziel, B.~D.} \emph{et~al.}
\newblock \bibinfo{journal}{\bibinfo{title}{Urbanization and humidity shape the
  intensity of influenza epidemics in {US} cities}}.
\newblock {\emph{\JournalTitle{Science}}} \textbf{\bibinfo{volume}{362}},
  \bibinfo{pages}{75--79} (\bibinfo{year}{2018}).

\bibitem{lee2020epidemic}
\bibinfo{author}{Lee, V.~J.} \emph{et~al.}
\newblock \bibinfo{journal}{\bibinfo{title}{Epidemic preparedness in urban
  settings: new challenges and opportunities}}.
\newblock {\emph{\JournalTitle{The Lancet Infectious Diseases}}}
  \textbf{\bibinfo{volume}{20}}, \bibinfo{pages}{527--529}
  (\bibinfo{year}{2020}).

\bibitem{bassolas2019}
\bibinfo{author}{Bassolas, A.} \emph{et~al.}
\newblock \bibinfo{journal}{\bibinfo{title}{Hierarchical organization of urban
  mobility and its connection with city livability}}.
\newblock {\emph{\JournalTitle{Nature Communications}}}
  \textbf{\bibinfo{volume}{10}}, \bibinfo{pages}{4817} (\bibinfo{year}{2019}).

\bibitem{48778}
\bibinfo{author}{Wilson, R.} \emph{et~al.}
\newblock \bibinfo{title}{Differentially private sql with bounded user
  contribution} (\bibinfo{year}{2020}).

\bibitem{louail2014}
\bibinfo{author}{Louail, T.} \emph{et~al.}
\newblock \bibinfo{journal}{\bibinfo{title}{From mobile phone data to the
  spatial structure of cities}}.
\newblock {\emph{\JournalTitle{Scientific Reports}}}
  \textbf{\bibinfo{volume}{4}}, \bibinfo{pages}{5276} (\bibinfo{year}{2014}).

\bibitem{ewing2015}
\bibinfo{author}{Ewing, R.} \& \bibinfo{author}{Hamidi, S.}
\newblock \bibinfo{journal}{\bibinfo{title}{Compactness versus sprawl: A review
  of recent evidence from the united states}}.
\newblock {\emph{\JournalTitle{Journal of Planning Literature}}}
  \textbf{\bibinfo{volume}{30}}, \bibinfo{pages}{413--432}
  (\bibinfo{year}{2015}).

\bibitem{bettencourt2008real}
\bibinfo{author}{Bettencourt, L.~M.} \& \bibinfo{author}{Ribeiro, R.~M.}
\newblock \bibinfo{journal}{\bibinfo{title}{Real time bayesian estimation of
  the epidemic potential of emerging infectious diseases}}.
\newblock {\emph{\JournalTitle{PLOS ONE}}} \textbf{\bibinfo{volume}{3}}
  (\bibinfo{year}{2008}).

\bibitem{tizzoni2020}
\bibinfo{author}{Tizzoni, M.}
\newblock \bibinfo{title}{Estimating {COVID-19's Rt in Real-Time}}.
\newblock
  \bibinfo{howpublished}{\url{https://github.com/micheletizzoni/covid-19}}.
\newblock \bibinfo{note}{Accessed in July 2020}.

\bibitem{balcan2010}
\bibinfo{author}{Balcan, D.} \emph{et~al.}
\newblock \bibinfo{journal}{\bibinfo{title}{Modeling the spatial spread of
  infectious diseases: The global epidemic and mobility computational model}}.
\newblock {\emph{\JournalTitle{Journal of Computational Science}}}
  \textbf{\bibinfo{volume}{1}}, \bibinfo{pages}{132--145}
  (\bibinfo{year}{2010}).

\bibitem{domenico2020}
\bibinfo{author}{Di~Domenico, L.}, \bibinfo{author}{Pullano, G.},
  \bibinfo{author}{Sabbatini, C.~E.}, \bibinfo{author}{Bo{\"e}lle, P.-Y.} \&
  \bibinfo{author}{Colizza, V.}
\newblock \bibinfo{journal}{\bibinfo{title}{Impact of lockdown on {COVID-19}
  epidemic in {\^i}le-de-france and possible exit strategies}}.
\newblock {\emph{\JournalTitle{BMC Medicine}}} \textbf{\bibinfo{volume}{18}},
  \bibinfo{pages}{1--13} (\bibinfo{year}{2020}).

\bibitem{sattespiel1995}
\bibinfo{author}{Sattenspiel, L.} \& \bibinfo{author}{Dietz, K.}
\newblock \bibinfo{journal}{\bibinfo{title}{A structured epidemic model
  incorporating geographic mobility among regions}}.
\newblock {\emph{\JournalTitle{Mathematical Biosciences}}}
  \textbf{\bibinfo{volume}{128}}, \bibinfo{pages}{71 -- 91},
  \doiprefix\url{https://doi.org/10.1016/0025-5564(94)00068-B}
  (\bibinfo{year}{1995}).

\bibitem{aleta2020}
\bibinfo{author}{Aleta, A.} \emph{et~al.}
\newblock \bibinfo{journal}{\bibinfo{title}{Modelling the impact of testing,
  contact tracing and household quarantine on second waves of {COVID-19}}}.
\newblock {\emph{\JournalTitle{Nature Human Behaviour}}}
  \textbf{\bibinfo{volume}{4}}, \bibinfo{pages}{964--971}
  (\bibinfo{year}{2020}).

\bibitem{arenas2020}
\bibinfo{author}{Arenas, A.} \emph{et~al.}
\newblock \bibinfo{journal}{\bibinfo{title}{Modeling the spatiotemporal
  epidemic spreading of {COVID-19} and the impact of mobility and social
  distancing interventions}}.
\newblock {\emph{\JournalTitle{Physical Review X}}}
  \textbf{\bibinfo{volume}{10}}, \bibinfo{pages}{041055}
  (\bibinfo{year}{2020}).

\bibitem{colizza2020}
\bibinfo{author}{Pullano, G.}, \bibinfo{author}{Valdano, E.},
  \bibinfo{author}{Scarpa, N.}, \bibinfo{author}{Rubrichi, S.} \&
  \bibinfo{author}{Colizza, V.}
\newblock \bibinfo{journal}{\bibinfo{title}{Evaluating the effect of
  demographic factors, socioeconomic factors, and risk aversion on mobility
  during the {COVID-19} epidemic in {France} under lockdown: a population-based
  study}}.
\newblock {\emph{\JournalTitle{The Lancet Digital Health}}}
  \textbf{\bibinfo{volume}{2}}, \bibinfo{pages}{e638--e649}
  (\bibinfo{year}{2020}).

\bibitem{starnini2020}
\bibinfo{author}{Starnini, M.}, \bibinfo{author}{Aleta, A.},
  \bibinfo{author}{Tizzoni, M.} \& \bibinfo{author}{Moreno, Y.}
\newblock \bibinfo{journal}{\bibinfo{title}{Impact of the accuracy of
  case-based surveillance data on the estimation of time-varying reproduction
  numbers}}.
\newblock {\emph{\JournalTitle{medRxiv}}}  (\bibinfo{year}{2020}).

\bibitem{ISSitaly}
\bibinfo{title}{{Istituto Superiore di Sanit\'a. COVID-19 integrated
  surveillance data in Italy}} (\bibinfo{year}{2020}).

\bibitem{fraiberger2020}
\bibinfo{author}{Fraiberger, S.~P.} \emph{et~al.}
\newblock \bibinfo{journal}{\bibinfo{title}{Uncovering socioeconomic gaps in
  mobility reduction during the {COVID-19} pandemic using location data}}.
\newblock {\emph{\JournalTitle{arXiv preprint arXiv:2006.15195}}}
  (\bibinfo{year}{2020}).

\end{thebibliography}

\end{document}